\UseRawInputEncoding
\documentclass[pra,twocolumn,shownopacs,amssymb]{revtex4}
\usepackage{graphicx,psfrag,amsmath,amssymb}
    
\begin{document}

\title{Quantum-geometric perspective on spin-orbit-coupled Bose superfluids}

\author{A. L. Suba{\c s}{\i}$^{1}$ and M. Iskin$^{2}$}

\affiliation{ 
$^{1}$Department of Physics, Faculty of Science and Letters, Istanbul
Technical University, 34469 Maslak, Istanbul, Turkey. \\
$^{2}$Department of Physics, Ko{\c c} University, Rumelifeneri Yolu,
34450 Sar{\i}yer, Istanbul, Turkey. 
}

\date{\today}

\begin{abstract}

We employ the Bogoliubov approximation to study how the quantum geometry 
of the helicity states affects the superfluid properties of a spin-orbit-coupled 
Bose gas in continuum. In particular we derive the low-energy Bogoliubov 
spectrum for a plane-wave condensate in the lower helicity band and show that 
the geometric contributions to the sound velocity are distinguished by their linear 
dependences on the interaction strength, i.e., they are in sharp contrast to the 
conventional contribution which has a square-root dependence. We also discuss
the roton instability of the plane-wave condensate against the stripe phase and 
determine their phase transition boundary. In addition we derive the superfluid 
density tensor by imposing a phase-twist on the condensate order parameter and 
study the relative importance of its contribution from the interband processes that 
is related to the quantum geometry.

\end{abstract}

\maketitle

\section{Introduction}
\label{sec:intro}

Recent studies have shown that the quantum geometry of the Bloch states can 
play important roles in characterizing some of the fundamental properties of Fermi 
superfluids (SFs)~\cite{peotta15, liang17}. The physical mechanism is quite clear 
in a multiband lattice: the geometric effects originate from the dressing of the 
effective mass of the SF carriers by the interband processes, which in return 
controls those SF properties that depend on the carrier mass. Besides the SF density/weight, 
the list includes the velocity of the low-energy Goldstone modes and the critical BKT 
temperature~\cite{peotta15, liang17, iskin18a, iskin18b, iskin20a, iskin20b, julku18, wang20}. 
On the other hand the intraband processes give rise to the conventional effects.
Depending on the band structure and the strength of the interparticle interactions, 
it has been established that the geometric effects can become sizeable and may 
even dominate in an isolated flat band~\cite{peotta15}. Furthermore such geometric 
effects on Fermi SFs can be traced all the way back to the two-body problem 
in a multiband lattice in vacuum~\cite{torma18, iskin21}.

Despite the growing number of recent works exposing the role of quantum geometry 
for the Fermi SFs, there is a lack of understanding in the bosonic counterparts 
which are much less studied~\cite{julku21a, julku21b, yangnote}. For instance Julku et al. have 
considered a weakly-interacting BEC in a flat band, and showed that the speed of 
sound has a linear dependence on the interaction strength and a square-root 
dependence on the quantum metric of the condensed Bloch state~\cite{julku21a, julku21b}. 
They have also showed that the quantum depletion is dictated solely by the quantum 
geometry and the SF weight has a quantum-geometric origin. 

Motivated by the success of analogous works on spin-orbit-coupled Fermi 
SFs~\cite{iskin18a, iskin18b, iskin20a, julku18}, here we investigate the SF 
properties of a spin-orbit-coupled Bose gas from a quantum-geometric perspective. 
Our work differs from the existing literature in several ways~\cite{li15, zhai15, zhang16a}.
In particular we derive the low-energy Bogoliubov spectrum for a plane-wave condensate 
in the lower helicity band and identify the geometric contributions to the sound velocity. 
The geometric effects survive only when the single-particle Hamiltonian has a 
a $\sigma_z$ term in the pseudospin basis that is coupled with a $\sigma_x$ (and/or 
equivalently a $\sigma_y$) term.
In contrast to the conventional contribution that has a square-root dependence on the 
interaction strength, we find that the geometric ones are distinguished by a linear 
dependence. Similar to the Fermion problem where the geometric effects dress 
the effective mass of the Goldstone modes, here one can also interpret the 
geometric terms in terms of a dressed effective mass for the Bogoliubov modes.
We also discuss the roton instability of the plane-wave ground state 
against the stripe phase and determine the phase transition boundary. 
All of these results are achieved analytically by reducing the $4 \times 4$ Bogoliubov
Hamiltonian (that involves both lower and upper helicity bands) down to $2 \times 2$ 
through projecting the system onto the lower helicity band. The projected Hamiltonian 
works extremely well except for a tiny region in momentum-space around the point 
where the helicity bands are degenerate.
In addition we derive the SF density tensor by imposing a phase-twist on 
the condensate order parameter and analyze the relative importance of its contribution 
from the interband processes~\cite{yangnote}. 

The rest of the paper is organized as follows. We begin with the theoretical model
in Sec.~\ref{sec:tm}: the many-body Hamiltonian is introduced in Sec.~\ref{sec:ham} 
and the noninteracting helicity spectrum is reviewed in Sec.~\ref{sec:hb}.
Then we present the Bogoliubov mean-field theory for a plane-wave condensate 
in Sec.~\ref{sec:bt}: the four branches of the full Bogoliubov spectrum are discussed 
in Sec.~\ref{sec:bs} and the two branches of the projected (i.e., to the lower-helicity 
band) Bogoliubov spectrum are derived in Sec.~\ref{sec:ps}. 
Furthermore, by analyzing the resultant Bogoliubov spectrum in the low-energy regime, 
we find closed-form analytic expressions for the Bogoliubov modes in Sec.~\ref{sec:lowq} 
and for the roton instability of the plane-wave condensate against the stripe phase 
in Sec.~\ref{sec:sgs}. Finally we derive and analyze the SF density tensor 
and condensate density in Sec.~\ref{sec:scd}.
The paper ends with a summary of our conclusions given in Sec.~\ref{sec:conc}.

\section{Theoretical Model}
\label{sec:tm}

In order to study the interplay between a BEC and SOC, and having cold-atom
systems in mind, here we consider a two-component atomic Bose gas that is 
characterized by a weakly-repulsive zero-ranged (contact) interactions in continuum.
It is customary to refer to such a two-component bosonic system as the 
pseudospin-$1/2$ Bose gas.

\subsection{Pseudospin-$1/2$ Bose Gas}
\label{sec:ham}

In particular, by making use of the momentum-space representation, we express the 
single-particle Hamiltonian in the usual form
\begin{align}
\mathcal{H}_0 = \sum_\mathbf{k} \Lambda_\mathbf{k}^\dagger
\left[ \big(\varepsilon_\mathbf{k} + \varepsilon_\mathbf{k_0}\big) \sigma_0
+ \frac{\mathbf{d}_\mathbf{k} \cdot \boldsymbol{\sigma}}{m} \right]
\Lambda_\mathbf{k},
\label{eqn:H0}
\end{align}
where
$
\mathbf{k} = (k_x, k_y, k_z)
$
is the momentum vector with $\hbar = 1$ and
$
\Lambda_\mathbf{k}^\dagger = \big( a_{\uparrow \mathbf{k}}^\dagger 
\; a_{\downarrow \mathbf{k}}^\dagger \big)
$
is a two-component spinor with the creation operator 
$
a_{\sigma \mathbf{k}}^\dagger
$
for a pseudospin-$\sigma$ particle in state
$
| \sigma \mathbf{k} \rangle = a_{\sigma \mathbf{k}}^\dagger | 0 \rangle.
$
Here $\sigma = \{\uparrow, \downarrow\}$ labels the two components of the
Bose gas and $| 0 \rangle$ is the vacuum state. The first term
$
\varepsilon_\mathbf{k} = k^2/(2m)
$
is the kinetic energy of a particle where $\varepsilon_\mathbf{k_0}$ is a convenient 
choice of an energy offset ($\mathbf{k_0}$ is defined below) and $\sigma_0$ is an 
identity matrix. 
The second term is the so-called SOC where
$
\boldsymbol{\sigma} = (\sigma_x, \sigma_y, \sigma_z) 
$
is a vector of Pauli spin matrices and
$
\mathbf{d}_\mathbf{k} = \big(d^x_\mathbf{k}, d^y_\mathbf{k}, d^z_\mathbf{k}\big)
$
is the SOC field with linearly dispersing components
$
d^i_\mathbf{k} = \alpha_i k_i.
$
Here we choose $\alpha_i \ge 0$ and $\alpha_x \ge \{\alpha_y, \alpha_z\}$ without 
the loss of generality.

Similarly a compact way to express the intraspin and interspin interaction terms is
\begin{align}
\mathcal{H}_U = \frac{1}{2V} \sum_{\substack{\sigma\sigma' \\ \mathbf{k_1}+\mathbf{k_2} = \mathbf{k_3}+\mathbf{k_4}}}
U_{\sigma\sigma'} a_{\sigma \mathbf{k_1}}^\dagger a_{\sigma' \mathbf{k_2}}^\dagger
a_{\sigma' \mathbf{k_3}} a_{\sigma \mathbf{k_4}},
\label{eqn:HU}
\end{align}
where $V$ is the volume and $U_{\sigma\sigma'} \ge 0$ is the strength of the interactions.
Here we consider a sufficiently weak $U_{\uparrow\downarrow}$ in order to prevent competing 
phases that are beyond the scope of this paper. 
See Sec.~\ref{sec:sgs} for a detailed account of the stability analysis.
In addition we include a chemical potential term
$
\mathcal{H}_\mu = -\sum_{\sigma \mathbf{k}} \mu_\sigma a_{\sigma \mathbf{k}}^\dagger a_{\sigma \mathbf{k}}
$
to the total Hamiltonian $\mathcal{H} = \mathcal{H}_0 + \mathcal{H}_U + \mathcal{H}_\mu$ 
of the system, and determine $\mu_\sigma$ in a self-consistent fashion.

\subsection{Helicity Bands}
\label{sec:hb}

Let us first discuss the single-particle ground state. The eigenvalues of the Hamiltonian
matrix shown in Eq.~(\ref{eqn:H0}) can be written as
\begin{align}
\xi_{s\mathbf{k}} = \varepsilon_\mathbf{k} + \varepsilon_\mathbf{k_0} + s \frac{d_\mathbf{k}}{m},
\label{eqn:helicitybands}
\end{align}
where $s = \pm$ labels, respectively, the upper and lower band and 
$d_\mathbf{k} = |\mathbf{d}_\mathbf{k}|$ is the magnitude of the SOC field.
Therefore the single-particle (helicity) spectrum exhibits two branches due to SOC.
In the pseudospin basis $| \sigma \mathbf{k} \rangle$, the corresponding eigenvectors 
(i.e., helicity basis) 
$
| s \mathbf{k} \rangle = a_{s\mathbf{k}}^\dagger |0\rangle
$
can be represented as
$
| +, \mathbf{k} \rangle = 
\begin{pmatrix}
u_\mathbf{k} & v_\mathbf{k}e^{i\varphi_\mathbf{k}}
\end{pmatrix}^\mathrm{T}
$
for the upper and
$
| -, \mathbf{k} \rangle = 
\begin{pmatrix}
-v_\mathbf{k}e^{-i\varphi_\mathbf{k}} & u_\mathbf{k}
\end{pmatrix}^\mathrm{T}
$
for the lower helicity band, where
$
u_\mathbf{k} = \sqrt{(d_\mathbf{k}+d^z_\mathbf{k})/(2d_\mathbf{k})},
$
$
v_\mathbf{k} = \sqrt{(d_\mathbf{k}-d^z_\mathbf{k})/(2d_\mathbf{k})},
$
$
\varphi_\mathbf{k} = \arg(d^x_\mathbf{k} + i d^y_\mathbf{k}),
$
and $\mathrm{T}$ denotes the transpose. Alternatively, 
\begin{align}
\begin{pmatrix}
a_{\uparrow \mathbf{k}} \\ a_{\downarrow \mathbf{k}}  
\end{pmatrix}
= 
\begin{pmatrix}
u_\mathbf{k} & -v_\mathbf{k} e^{-i\varphi_\mathbf{k}} \\
v_\mathbf{k} e^{i\varphi_\mathbf{k}} & u_\mathbf{k} 
\end{pmatrix}
\begin{pmatrix}
a_{+, \mathbf{k}} \\ a_{-, \mathbf{k}} 
\end{pmatrix} \nonumber
\end{align}
is the transformation between the annihilation operators for the pseudospin 
and helicity states.

\begin{figure}[!ht]
\centering
\resizebox{1.0\columnwidth}{!}{%
\includegraphics{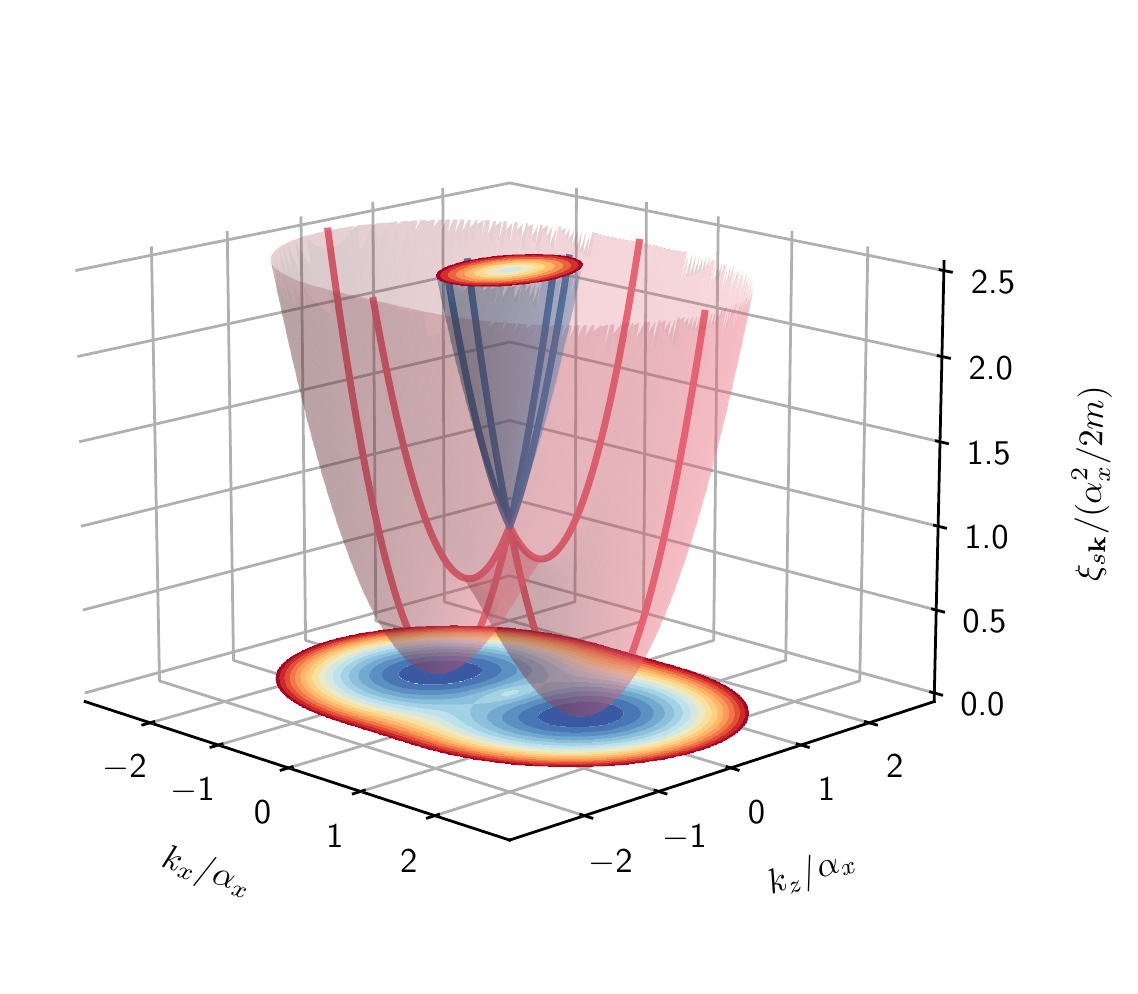}
}
\caption{
Helicity bands $\xi_{s \mathbf{k}}$ ( in units of $\alpha_x^2/2m$) are shown for 
$\alpha_x = 2\alpha_z$ and $\alpha_y=0$ at $k_y = 0$.
The upper (red) and lower (blue) bands touch at $\mathbf{k}=\mathbf{0}$.
The single-particle ground state is doubly degenerate at $\mathbf{k}=(\pm \alpha_x,0,0)$.
\label{fig:hb}
}
\end{figure}

For notational convenience, the lower helicity state $| -, \mathbf{k} \rangle$ is
denoted as $| \phi_\mathbf{k} \rangle$ in the rest of the paper. Then the 
single-particle ground state $|\phi_\mathbf{k_0} \rangle$ is determined by 
setting
$
\partial \xi_{-, \mathbf{k}}/\partial k_i = 0,
$ 
leading to either $k_i = 0$ or $\alpha_i^2 = d_\mathbf{k}$. Here we choose
$\mathbf{k_0} = (\alpha_x, 0, 0)$ without the loss of generality~\cite{cui13, ozawa12, baym14},
for which case the single-particle ground-state energy $\xi_{-,\mathbf{k_0}} = 0$ 
vanishes (see Fig.~\ref{fig:hb}) and the single-particle ground state
$
| \phi_\mathbf{k_0} \rangle = 
\begin{pmatrix}
-1/\sqrt{2} & 1/\sqrt{2}
\end{pmatrix}^\mathrm{T}
$
admits a real representation. 
Note that the ground state is at least two-fold degenerate with the opposite-momentum
state 
$
| \phi_{-\mathbf{k_0}} \rangle = 
\begin{pmatrix}
1/\sqrt{2} & 1/\sqrt{2}
\end{pmatrix}^\mathrm{T},
$
and we highlight its competing role in Sec.~\ref{sec:sgs}.
Having introduced the theoretical model, and discussed its single-particle ground state, 
next we analyze the many-body ground state within the Bogoliubov mean-field approximation.

\section{Bogoliubov Theory}
\label{sec:bt}

Under the Bogoliubov mean-field approximation, the many-body ground state is known to be 
either a plane-wave condensate or a stripe phase depending on the relative strengths between 
the intraspin and interspin interactions~\cite{cui13, ozawa12, baym14, wang10, barnett12}. 
See Sec.~\ref{sec:sgs} for a detailed account of the stability analysis.
Assuming that $U_{\uparrow\downarrow}$ is sufficiently weak, here we concentrate only 
on the former phase.

\subsection{Bogoliubov Spectrum}
\label{sec:bs}

In order to describe the many-body ground state $| \phi_\mathbf{k_0} \rangle$ that is 
macroscopically occupied by $N_0$ particles, we replace the annihilation and creation 
operators in accordance with
$
a_{\sigma \mathbf{k}} = \Delta_\sigma \sqrt{V} \delta_{\mathbf{k} \mathbf{k_0}} 
+ \tilde{a}_{\sigma \mathbf{k}}.
$
Here the complex field
$
\Delta_\sigma = \sqrt{n_0} \langle \sigma | \phi_\mathbf{k_0} \rangle
$
corresponds to the mean-field order parameter for the condensate with condensate 
density $n_0 = N_0/V$, $\delta_{ij}$ is a Kronecker-delta, and the operator 
$\tilde{a}_{\sigma \mathbf{k}}$ denotes the fluctuations on top of the ground state. 
Following the usual recipe, we neglect the third- and fourth-order fluctuation terms
in the interaction Hamiltonian. Then the excitations are described by the so-called
Bogoliubov Hamiltonian
\begin{align} 
\mathcal{H}_\mathrm{B} &= \frac{1}{2} \sum_\mathbf{q}^\prime \Psi_\mathbf{q}^\dagger 
\begin{pmatrix}
\mathbf{H}^{pp}_\mathbf{q} & \mathbf{H}^{ph}_\mathbf{q} \\
\mathbf{H}^{hp}_\mathbf{q} & \mathbf{H}^{hh}_\mathbf{q} 
\end{pmatrix}
\Psi_\mathbf{q},
\label{eqn:HB}
\\
\mathbf{H}^{pp}_\mathbf{q} &= 
\begin{pmatrix}
K_{\uparrow\mathbf{q}}
& U_{\uparrow\downarrow}\Delta_\uparrow \Delta_\downarrow^* \\
U_{\uparrow\downarrow}\Delta_\uparrow^* \Delta_\downarrow & 
K_{\downarrow\mathbf{q}}
\end{pmatrix}
+ \frac{\mathbf{d}_\mathbf{k_0+q} \cdot \boldsymbol{\sigma}}{m},
\\
\mathbf{H}^{ph}_\mathbf{q} &= 
\begin{pmatrix}
U_{\uparrow\uparrow}\Delta_\uparrow^2 & 
U_{\uparrow\downarrow}\Delta_\uparrow \Delta_\downarrow \\
U_{\uparrow\downarrow}\Delta_\uparrow \Delta_\downarrow & 
U_{\downarrow\downarrow}\Delta_\downarrow^2
\end{pmatrix},
\end{align}
where
$
\Psi_\mathbf{q}^\dagger = 
\big(
\tilde{a}_{\uparrow, \mathbf{k_0+q}}^\dagger \;
\tilde{a}_{\downarrow, \mathbf{k_0+q}}^\dagger \;
\tilde{a}_{\uparrow, \mathbf{k_0-q}} \;
\tilde{a}_{\downarrow, \mathbf{k_0-q}}
\big)
$
is a four-component spinor and
$
K_{\sigma\mathbf{q}} = \varepsilon_\mathbf{k_0+q} + \varepsilon_\mathbf{k_0} - \mu_\sigma 
+ 2U_{\sigma\sigma}|\Delta_\sigma|^2 + U_{\uparrow\downarrow}|\Delta_{-\sigma}|^2
$
with the index $-\sigma$ denoting the opposite component of the spin. 
The other terms are simply related via
$
\mathbf{H}^{hh}_\mathbf{q} = (\mathbf{H}^{pp}_{-\mathbf{q}})^* 
$
and
$
\mathbf{H}^{hp}_\mathbf{q} = (\mathbf{H}^{ph}_\mathbf{q})^\dagger.
$
The prime symbol indicates that the summation is over all of the non-condensed states.
In this approximation, $\mu_\sigma$ is determined by setting the first-order fluctuation 
terms to $0$, leading to
$
\mu_\sigma = U_{\sigma\sigma} |\Delta_\sigma|^2 + U_{\uparrow\downarrow} |\Delta_{-\sigma}|^2.
$
Note that
$
\Delta_\uparrow = -\Delta_\downarrow = -\sqrt{n_0/2}
$
are real for our particular choice for the ground state $| \phi_\mathbf{k_0} \rangle$.

The Bogoliubov spectrum $E_{s\mathbf{q}}^n$ is determined by the eigenvalues of 
$\tau_z \mathbf{H}_\mathbf{q}$~\cite{julku21a, julku21b}, i.e.,
\begin{align}
\tau_z \mathbf{H}_\mathbf{q} | \chi_{s \mathbf{q}}^n \rangle 
= E_{s\mathbf{q}}^n  | \chi_{s \mathbf{q}}^n \rangle,
\label{eqn:bogmod}
\end{align}
where $\tau_z$ is a Pauli matrix acting only on the particle-hole sector, 
$\mathbf{H}_\mathbf{q}$ is the $4\times4$ Hamiltonian matrix shown in Eq.~(\ref{eqn:HB}),
and $| \chi_{s \mathbf{q}}^n \rangle$ is the corresponding Bogoliubov state. 
Here $n = \pm$ labels, respectively, the upper and lower Bogoliubov band, 
and $s = \pm$ labels, respectively, the quasiparticle and quasihole branch for a
given band $n$, leading to four Bogoliubov modes for a given $\mathbf{q}$.
The Bogoliubov states are normalized in the usual way, i.e., if we denote 
$
| \chi_{s \mathbf{q}}^n \rangle = 
\begin{pmatrix}
| \chi_{s \mathbf{q}}^n \rangle_1 \\
| \chi_{s \mathbf{q}}^n \rangle_2
\end{pmatrix}
$
then
$
{_1\langle} \chi_{s \mathbf{q}}^n | \chi_{s \mathbf{q}}^n \rangle_1 -
{_2\langle} \chi_{s \mathbf{q}}^n | \chi_{s \mathbf{q}}^n \rangle_2 = s.
$
While the Bogoliubov spectrum exhibits
$
E_{s\mathbf{q}}^n = -E_{-s,-\mathbf{q}}^n
$
as a manifestation of the quasiparticle-quasihole symmetry, Eq.~(\ref{eqn:bogmod}) 
does not allow for a closed-form analytic solution in general, and its characterization 
requires a fully numerical procedure.

In order to gain some analytical insight into the low-energy Bogoliubov modes, 
we assume that the energy gap between the lower and upper helicity bands 
nearby the ground state $| \phi_\mathbf{k_0} \rangle$ is much larger than the interaction 
energy. This occurs when the SOC energy scale is much stronger than the interaction 
energy scale. In this case the occupation of the upper band is negligible, and the system 
can be projected solely to the lower band as discussed next.

\subsection{Projected System}
\label{sec:ps}

The total Hamiltonian $\mathcal{H}$ of the system can be projected to the lower 
helicity band as follows~\cite{cui13}. Using the identity operator 
$
\sigma_0 = \sum_s | s\mathbf{k} \rangle \langle s\mathbf{k} |
$
for a given $\mathbf{k}$, we first reexpress
$
a_{\sigma \mathbf{k}} = \sum_s \langle \sigma | s \mathbf{k} \rangle a_{s \mathbf{k}},
$
and discard those terms that involve the upper band, i.e.,
$
a_{\sigma \mathbf{k}} \to \langle \sigma | \phi_\mathbf{k} \rangle a_{-, \mathbf{k}}.
$
This procedure leads to
\begin{align}
&h_0 + h_\mu = \sum_\mathbf{k} \big( \xi_{-,\mathbf{k}} - \mu \big) a_{-, \mathbf{k}}^\dagger  a_{-, \mathbf{k}}, \\
&h_U = \frac{1}{2V} \sum_{\mathbf{k_1}+\mathbf{k_2} = \mathbf{k_3}+\mathbf{k_4}}
f_{\mathbf{k_1} \mathbf{k_2}}^{\mathbf{k_3} \mathbf{k_4}} 
a_{-, \mathbf{k_1}}^\dagger a_{-, \mathbf{k_2}}^\dagger
a_{-, \mathbf{k_3}} a_{-, \mathbf{k_4}}, \\
&f_{\mathbf{k_1} \mathbf{k_2}}^{\mathbf{k_3} \mathbf{k_4}} 
= \sum_{\sigma\sigma'} U_{\sigma\sigma'} 
\langle \phi_\mathbf{k_1}|\sigma \rangle
\langle \phi_\mathbf{k_2}|\sigma' \rangle
\langle \sigma' | \phi_\mathbf{k_3} \rangle
\langle \sigma | \phi_\mathbf{k_4} \rangle,
\label{eqn:fk}
\end{align}
where
$
\mu = (\mu_\uparrow + \mu_\downarrow)/2
$
is the effective chemical potential and
$
f_{\mathbf{k_1} \mathbf{k_2}}^{\mathbf{k_3} \mathbf{k_4}} 
= U_{\uparrow\uparrow} v_\mathbf{k_1} v_\mathbf{k_2} v_\mathbf{k_3} v_\mathbf{k_4}
e^{i(\varphi_\mathbf{k_1}+\varphi_\mathbf{k_2}-\varphi_\mathbf{k_3}-\varphi_\mathbf{k_4})}
+ U_{\downarrow\downarrow} u_\mathbf{k_1} u_\mathbf{k_2} u_\mathbf{k_3} u_\mathbf{k_4}
+U_{\uparrow\downarrow} v_\mathbf{k_1} u_\mathbf{k_2} u_\mathbf{k_3} v_\mathbf{k_4}
e^{i(\varphi_\mathbf{k_1}-\varphi_\mathbf{k_4})}
$
is the effective long-range interaction for the projected system. We note that the long-range 
nature of the effective interaction plays a crucial role in the Bogoliubov spectrum as 
discussed in Sec.~\ref{sec:sgs}. 

\begin{figure}[!ht]
\centering
\resizebox{1.0\columnwidth}{!}{%
\includegraphics{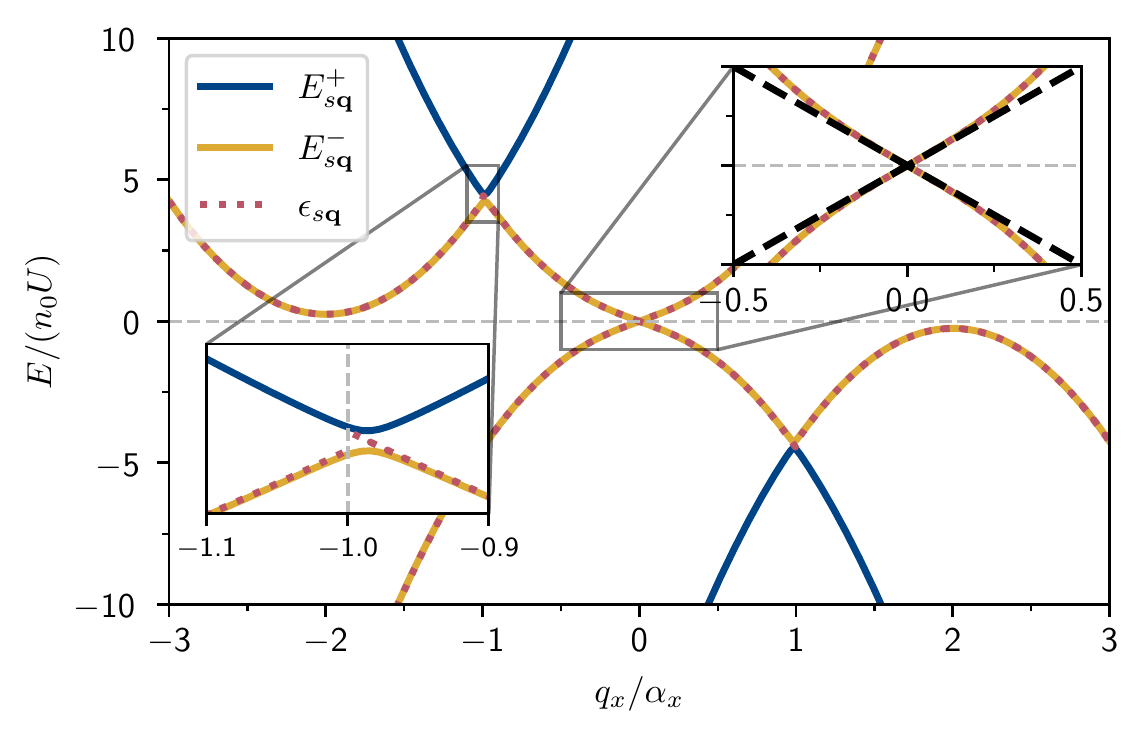}
}
\caption{
Bogoliubov spectrum is shown as a function of $q_x$ when $q_y=0=q_z$,
$
U = U_{\uparrow\uparrow}=2 U_{\downarrow\downarrow} = 4 U_{\uparrow\downarrow}
$, 
$\alpha_x = \alpha_y = 2/\xi$ with the healing length $\xi = 1/\sqrt{2m n U}$, 
and $\alpha_z=0$. Here the total particle density $n \approx n_0$ is set to $n a^3=10^{-6}$ 
where $a = mU/(4\pi)$ is the scattering length.
The full spectrum (solid lines) is shown together with the projected one (dotted lines) that is
given by Eq.~(\ref{eqn:esq}). In addition the low-$q$ expansion Eq.~(\ref{eqn:elow}) is 
shown as dashed black lines in the right inset. 
If one sets $U_{\uparrow\uparrow}= U_{\downarrow\downarrow}$ then the band gap 
shown in the left inset disappears, i.e., see Sec.~\ref{sec:jump} for the analysis of the
spurious jumps at $q_x = \mp \alpha_x$.
If one sets 
$
U_{\uparrow\downarrow} = U_{\uparrow\uparrow}=U_{\downarrow\downarrow}
$ 
then two additional zero-energy modes appear at $q_x = \mp 2\alpha_x$, i.e., 
see Sec.~\ref{sec:sgs} for the analysis of the roton instability.
\label{fig:fig1}
}
\end{figure}

Under the Bogoliubov mean-field approximation that is used in Sec.~\ref{sec:bs}, we replace 
the creation and annihilation operators in accordance with
$
a_{-, \mathbf{k}} = \sqrt{N_0} \delta_{\mathbf{k} \mathbf{k_0}} + \tilde{a}_{-, \mathbf{k}}
$
and set the first-order fluctuation terms to $0$. This leads to
$
\mu = n_0 f_{\mathbf{k_0} \mathbf{k_0}}^{\mathbf{k_0} \mathbf{k_0}}
= (n_0/4) \sum_{\sigma\sigma'} U_{\sigma\sigma'},
$
which is consistent with $\mu_\sigma$ that is found in Sec.~\ref{sec:bs}.
The zeroth-order fluctuation terms give 
$
-\mu N_0 + n_0 f_{\mathbf{k_0} \mathbf{k_0}}^{\mathbf{k_0} \mathbf{k_0}} N_0/2.
$
Then the excitations above the ground state are described by the Bogoliubov Hamiltonian
\begin{align} 
h_\mathrm{B} &= \frac{1}{2} \sum_\mathbf{q}^\prime \psi_\mathbf{q}^\dagger 
\begin{pmatrix}
h^{pp}_\mathbf{q} & h^{ph}_\mathbf{q} \\
h^{hp}_\mathbf{q} & h^{hh}_\mathbf{q} 
\end{pmatrix}
\psi_\mathbf{q},
\\
h^{pp}_\mathbf{q} &= \xi_{-,\mathbf{k_0+q}} - \mu + 
\frac{n_0}{2} \left(f_{\mathbf{k_0}, \mathbf{k_0+q}}^{\mathbf{k_0}, \mathbf{k_0+q}} 
+ f_{\mathbf{k_0+q}, \mathbf{k_0}}^{\mathbf{k_0+q}, \mathbf{k_0}} \right.\nonumber \\
&\left.\;\;\;\;\;\;\;\;\;\;\;\;\;\;\;\;\;\;\;\;\;\;\;
+ f_{\mathbf{k_0+q}, \mathbf{k_0}}^{\mathbf{k_0}, \mathbf{k_0+q}} 
+ f_{\mathbf{k_0}, \mathbf{k_0+q}}^{\mathbf{k_0+q}, \mathbf{k_0}} 
\right), 
\\
h^{ph}_\mathbf{q} &= \frac{n_0}{2}
\left(f_{\mathbf{k_0+q}, \mathbf{k_0-q}}^{\mathbf{k_0}, \mathbf{k_0}} 
+ f_{\mathbf{k_0-q}, \mathbf{k_0+q}}^{\mathbf{k_0}, \mathbf{k_0}} 
\right),
\end{align}
where
$
\psi_\mathbf{q}^\dagger = \big( 
\tilde{a}_{-, \mathbf{k_0+q}}^\dagger \; \tilde{a}_{-, \mathbf{k_0-q}} 
\big)
$
is a two-component spinor, and the other terms are simply related via
$
h^{hh}_\mathbf{q} = h^{pp}_{-\mathbf{q}}
$
and
$
h^{hp}_\mathbf{q}  = (h^{ph}_\mathbf{q})^*.
$
The Bogoliubov spectrum $\epsilon_{s\mathbf{q}}$ is determined by the eigenvalues 
of $\tau_z \mathbf{h}_\mathbf{q}$, leading to two Bogoliubov modes for a 
given $\mathbf{q}$, i.e.,
\begin{align}
\epsilon_{s \mathbf{q}} &= \frac{h^{pp}_\mathbf{q} - h^{hh}_\mathbf{q}}{2} + 
s \sqrt{\left(\frac{h^{pp}_\mathbf{q} + h^{hh}_\mathbf{q}}{2}\right)^2 - |h^{ph}_\mathbf{q}|^2},
\label{eqn:esq}
\\
\label{eqn:hpp}
h^{pp}_\mathbf{q} &= \xi_{-, \mathbf{k_0+q}} - \mu 
+ \frac{n_0}{2} \sum_{\sigma\sigma'} U_{\sigma\sigma'} 
|\langle \phi_\mathbf{k_0+q} | \sigma \rangle|^2  \nonumber \\
+ n_0 &\sum_{\sigma\sigma'} U_{\sigma\sigma'}
\langle \phi_\mathbf{k_0+q} | \sigma' \rangle 
\langle \sigma' | \phi_\mathbf{k_0} \rangle
\langle \phi_\mathbf{k_0} | \sigma \rangle
\langle \sigma | \phi_\mathbf{k_0+q} \rangle, \\
\label{eqn:hph}
h^{ph}_\mathbf{q} &= n_0 \sum_{\sigma\sigma'} U_{\sigma\sigma'}
\langle \phi_\mathbf{k_0+q} | \sigma \rangle 
\langle \sigma | \phi_\mathbf{k_0} \rangle
\langle \phi_\mathbf{k_0}^* | \sigma' \rangle
\langle \sigma' | \phi_\mathbf{k_0-q}^* \rangle.
\end{align}
Here $s = \pm$ labels, respectively, the quasiparticle and quasihole branch of the
lower Bogoliubov band (i.e., $n = -$) that is discussed in Sec.~\ref{sec:bs}.
See Fig.~\ref{fig:fig1} for their excellent numerical benchmark except for the
spurious jumps at $\mathbf{q} = \mp \mathbf{k_0}$ that are discussed in 
Sec.~\ref{sec:jump}. The Bogolibov spectrum exhibits
$
\epsilon_{+,\mathbf{q}} = -\epsilon_{-,-\mathbf{q}}
$
as a manifestation of the quasiparticle-quasihole symmetry.
Note that when $U_{\sigma\sigma'} = U \delta_{\sigma\sigma'}$, these expressions reduce 
exactly to those of Ref.~\cite{julku21a, julku21b} with $M = 2$, where our
$
h^{pp}_\mathbf{q} = \xi_{-, \mathbf{k_0+q}} + Un_0/2 
$
and $h^{ph}_\mathbf{q}$ correspond, respectively, to their $\mathbf{q}^2/(2m_{eff}) + \mu$
and $\mu \alpha(\mathbf{q})$ provided that $\mu = U n_0/2$ in this particular case.
Such a reduction may not be surprising since the intraspin interactions 
$U_{\uparrow\uparrow}$ and $U_{\downarrow\downarrow}$ play the roles of 
sublattice-dependent onsite interactions $U_{AA}$ and $U_{BB}$, and the interspin 
interaction $U_{\uparrow \downarrow}$ plays the role of a (long-range) inter-sublattice 
interaction $U_{AB}$. Thus our $U_{\sigma\sigma'} = U \delta_{\sigma\sigma'}$ limit 
corresponds precisely to the $U = U_{AA} = U_{BB}$ and $U_{AB} = 0$ case that is 
considered in Ref.~\cite{julku21a, julku21b}.

We can make further analytical progress through a low-$\mathbf{q}$ expansion around 
the ground state, and use the fact that
$
|\langle \sigma | \phi_\mathbf{k_0} \rangle|^2 = 1/2
$
for both pseudospin components, i.e., the $z$ component of $\mathbf{k_0}$ vanishes
for the ground state.

\subsection{Low-Momentum Expansion}
\label{sec:lowq}

Up to second order in $\mathbf{q}$, the low-energy expansions around 
the ground state $| \phi_\mathbf{k_0} \rangle$ can be written as
\begin{align}
h^{pp}_\mathbf{q} &= \frac{1}{2} \sum_{ij} q_i q_j M^{-1}_{ij} - \mu 
+ \frac{n_0}{2} \sum_{\sigma\sigma'} U_{\sigma\sigma'} 
+ 2n_0 \sum_{i \sigma\sigma'} q_i U_{\sigma\sigma'}  \nonumber \\
&\times \mathrm{Re} \langle \partial_i \phi_\mathbf{k} | \sigma \rangle
\langle \sigma | \phi_\mathbf{k_0} \rangle
+ n_0\sum_{ij \sigma\sigma'} q_i q_j U_{\sigma\sigma'} \nonumber \\
&\times \big(
\mathrm{Re} \langle \partial_i \partial_j \phi_\mathbf{k} | \sigma \rangle
\langle \sigma | \phi_\mathbf{k_0} \rangle 
+ \langle \partial_i \phi_\mathbf{k} | \sigma \rangle
\langle \sigma | \partial_j \phi_\mathbf{k} \rangle / 2 \nonumber \\
&+ \langle \partial_i \phi_\mathbf{k} | \sigma' \rangle
\langle \sigma' | \phi_\mathbf{k_0} \rangle
\langle \phi_\mathbf{k_0} | \sigma \rangle
\langle \sigma | \partial_j \phi_\mathbf{k} \rangle
\big),
\\
h^{ph}_\mathbf{q} &= \frac{n_0}{4} \sum_{\sigma\sigma'} U_{\sigma\sigma'}
+ \frac{n_0}{2} \sum_{ij\sigma\sigma'} q_i q_j U_{\sigma\sigma'} \big(
\langle \partial_i \partial_j \phi_\mathbf{k} | \sigma \rangle
\langle \sigma | \phi_\mathbf{k_0} \rangle  \nonumber \\
&-2\langle \partial_i \phi_\mathbf{k} | \sigma \rangle
\langle \sigma | \phi_\mathbf{k_0} \rangle
\langle \phi_\mathbf{k_0}^* | \sigma' \rangle
\langle \sigma' | \partial_j \phi_\mathbf{k}^* \rangle
\big),
\end{align}
where the spectrum of the lower helicity band is expanded as
$
\xi_{-, \mathbf{k_0+q}} = (1/2) \sum_{ij} q_i q_j M^{-1}_{ij}.
$
Here $\mathbf{M}^{-1}$ is the inverse of the effective-mass tensor
whose elements are given by $M^{-1}_{xx} = 1/m$, 
$M^{-1}_{yy} = 1/m - \alpha_y^2/(m\alpha_x^2)$,
$M^{-1}_{zz} = 1/m - \alpha_z^2/(m\alpha_x^2)$,
and 0 otherwise. In addition $\mathrm{Re}$ denotes the real part 
of an expression and $| \partial_i \phi_\mathbf{k} \rangle$ stands for 
$\partial |\phi_\mathbf{k} \rangle / \partial k_i$ in the $\mathbf{k \to k_0}$ limit. 
By plugging these expansions in Eq.~(\ref{eqn:esq}), and keeping up to 
second-order terms in $\mathbf{q}$, we obtain
\begin{align}
\label{eqn:esqe}
\epsilon_{s\mathbf{q}} &= 2n_0 \sum_{i \sigma\sigma'} q_i U_{\sigma\sigma'} 
\mathrm{Re} \langle \partial_i \phi_\mathbf{k} | \sigma \rangle
\langle \sigma | \phi_\mathbf{k_0} \rangle 
+ s \sqrt{X_\mathbf{q}}, \\
X_\mathbf{q} &= \mu \sum_{ij} q_i q_j \big[
 M^{-1}_{ij} + n_0 \sum_{\sigma\sigma'} U_{\sigma\sigma'} \big(
\langle \partial_i \phi_\mathbf{k} | \sigma \rangle
\langle \sigma | \partial_j \phi_\mathbf{k} \rangle \nonumber \\ 
+ &\mathrm{Re}  \langle \partial_i \partial_j \phi_\mathbf{k} | \sigma \rangle
\langle \sigma | \phi_\mathbf{k_0} \rangle
+ 2 \langle \partial_i \phi_\mathbf{k} | \sigma' \rangle
\langle \sigma' | \phi_\mathbf{k_0} \rangle
\langle \phi_\mathbf{k_0} | \sigma \rangle \nonumber \\
\times &\langle \sigma | \partial_j \phi_\mathbf{k} \rangle
+ 2 \mathrm{Re} \langle \partial_i \phi_\mathbf{k} | \sigma' \rangle
\langle \sigma' | \phi_\mathbf{k_0} \rangle
\langle \phi_\mathbf{k_0}^* | \sigma \rangle
\langle \sigma | \partial_j \phi_\mathbf{k}^* \rangle
\big) \big], \nonumber
\end{align}
for the low-energy Bogoliubov spectrum of the projected system. 
In addition to the conventional effective-mass term that depends only on the 
helicity spectrum, here we have the so-called geometric terms that depend 
also on the helicity states. The quantum geometry of the underlying Hilbert space 
is masked behind those terms that depend on $| \partial_i \phi_\mathbf{k} \rangle$ 
and $| \partial_i \partial_j \phi_\mathbf{k} \rangle$~\cite{julku21a, julku21b}. 
While most of these terms cancel one another, they lead to
\begin{align}
\epsilon_{s\mathbf{q}} &= n_0 \big( U_{\downarrow\downarrow}-U_{\uparrow\uparrow} \big) 
\frac{\alpha_z q_z}{2\alpha_x^2} +  \frac{s}{2} \sqrt{n_0(U_{\uparrow\uparrow}+U_{\downarrow\downarrow}+2U_{\uparrow\downarrow})} \nonumber \\
\times &\sqrt{\sum_{ij} q_i q_j M^{-1}_{ij} + 
n_0 \big(U_{\uparrow\uparrow}+U_{\downarrow\downarrow}-2U_{\uparrow\downarrow} \big) 
\frac{\alpha_z^2 q_z^2}{4 \alpha_x^4}},
\label{eqn:elow}
\end{align}
manifesting explicitly the quasiparticle-quasihole symmetry.
When $U_{\uparrow\uparrow} = U_{\downarrow\downarrow}$, Eq.~(\ref{eqn:elow}) is in full 
agreement with the recent literature for the reported parameters~\cite{cui13}. 
In addition see the right inset in Fig.~\ref{fig:fig1} for its numerical benchmark 
with Eq.~(\ref{eqn:esq}). 

Our work reveals that the linear term in $\alpha_z q_z$ that is outside 
of the square root as well as the quadratic term in $\alpha_z q_z$ that is in the inside 
have a quantum-geometric origin.
Note that the geometric terms that depend on $\alpha_x$ and $\alpha_y$ vanish all 
together. Thus we conclude that the geometric effects survive only in the presence of 
a finite $\sigma_z$ coupling assuming a $\sigma_x$ (and/or equivalently a $\sigma_y$) 
coupling to begin with. See Sec.~\ref{sec:hb} for our initial assumption in choosing 
$\mathbf{k_0}$. Although we choose a $\mathbf{k_0}$ that is symmetric in $y$ and $z$ 
directions, the condition $|\langle \sigma | \phi_\mathbf{k_0} \rangle|^2 = 1/2$ breaks 
this symmetry in general for other $\mathbf{k_0}$ values as it requires $k_{0z} = 0$.
The remaining geometric terms can be isolated from the conventional 
effective-mass term in the $\mathbf{q} \to (0, 0, q_z)$ limit when $\alpha_z \approx \alpha_x$,
leading to $q_i q_j M^{-1}_{ij} = 0$. Therefore this particular limit can be used to 
distinguish the geometric origin of sound velocity from the conventional one, i.e., unlike 
the conventional term that has $\propto \sqrt{U}$ dependence on the interaction 
strength, the geometric ones have $\propto U$ dependence. 
The square root vs. linear dependence is consistent with the recent results on 
multi-band Bloch systems~\cite{julku21a, julku21b}.
We note that the geometric term that is inside the square root can be incorporated into 
the conventional effective mass term, leading to a `dressed' effective mass
$
M^{-1}_{zz} \to M^{-1}_{zz} + n_0(U_{\uparrow\uparrow}+U_{\downarrow\downarrow}-2U_{\uparrow\downarrow}) \alpha_z^2/(4\alpha_x^4)
$
for the Bogoliubov modes~\cite{julku21a, julku21b}.
While this geometric dressing shares some similarities with the dressing of the 
effective-mass tensor of the Cooper pairs or the Goldstone modes in spin-orbit-coupled 
Fermi SFs, their mathematical structure is entirely different~\cite{iskin18b, iskin20a}. 
The latter involves a $\mathbf{k}$-space sum over the quantum-metric tensor 
of the helicity bands that is weighted by a function of other quantities including 
the excitation spectrum.

We note in passing that when $U_{\sigma\sigma'} = U \delta_{\sigma\sigma'}$, our 
Eq.~(\ref{eqn:esqe}) reduce exactly to that of Ref.~\cite{julku21a, julku21b} with $M = 2$, 
for which case we obtain
$
\epsilon_{s\mathbf{q}} = s \epsilon_\mathbf{q} 
$
with
$
\epsilon_\mathbf{q} = \big[(Un_0/2) \sum_{ij} q_i q_j \big(
 M^{-1}_{ij} + U n_0 \langle \partial_i \phi_\mathbf{k} | \partial_j \phi_\mathbf{k} \rangle
 + 2 U n_0 \sum_{\sigma} \mathrm{Re} \langle \partial_i \phi_\mathbf{k} | \sigma \rangle
\langle \sigma | \phi_\mathbf{k_0} \rangle
\langle \phi_\mathbf{k_0}^* | \sigma \rangle
\langle \sigma | \partial_j \phi_\mathbf{k}^* \rangle
\big) \big]^{1/2}.
$
Furthermore, using the fact that $|\phi_\mathbf{k_0} \rangle$ is real for the ground state, 
we find
$
\epsilon_\mathbf{q} = \big[(Un_0/2)\sum_{ij} q_i q_j \big(
 M^{-1}_{ij} + U n_0 \langle \partial_i \phi_\mathbf{k} | \partial_j \phi_\mathbf{k} \rangle
+ U n_0 \mathrm{Re} \langle \partial_i \phi_\mathbf{k} | \partial_j \phi_\mathbf{k}^* \rangle 
\big) \big]^{\frac{1}{2}}.
$
In comparison the quantum metric of the lower helicity band is defined by
$
g_{ij}^\mathbf{k} = \mathrm{Re} \langle \partial_i \phi_\mathbf{k} | 
\big( \sigma_0 - | \phi_\mathbf{k} \rangle \langle \phi_\mathbf{k} | \big)
|\partial_j \phi_\mathbf{k} \rangle,
$
and it reduces to
$
g_{ij}^\mathbf{k} = \langle \partial_i \phi_\mathbf{k} |\partial_j \phi_\mathbf{k} \rangle
$
only when $| \phi_\mathbf{k} \rangle$ is real for all $\mathbf{k}$.
This is because
$
\langle \partial_i \phi_\mathbf{k} | \phi_\mathbf{k}\rangle = - 
\langle \phi_\mathbf{k} | \partial_i \phi_\mathbf{k} \rangle = -
\langle \partial_i \phi_\mathbf{k}^* | \phi_\mathbf{k}^* \rangle
$ 
must vanish when $\phi_\mathbf{k}$ is real.
Thus we conclude that the geometric dressing of the effective mass of the Bogoliubov
modes can be written in terms of $g_{ij}^\mathbf{k}$ when $| \phi_\mathbf{k} \rangle$ 
is real for all $\mathbf{k}$. This is clearly the case when $d_\mathbf{k}^y = 0$ in
twoband lattices and when $\alpha_y = 0$ in spin-orbit-coupled Bose SFs.

Furthermore, when $\alpha_z \ne 0$, we find that the competition between the 
linear term in $q_z$ that is outside of the square root and the quadratic terms 
within the square root in Eq.~(\ref{eqn:elow}) causes an energetic instability 
(i.e., $\epsilon_{s\mathbf{q}}$ changes sign and becomes 
$\epsilon_{\pm,\mathbf{q} \lessgtr 0}$) in the $\mathbf{q} \to \mathbf{0}$ limit unless
\begin{align}
\frac{4U_{\uparrow\downarrow}^2 - (3U_{\uparrow\uparrow}-U_{\downarrow\downarrow})
(3U_{\downarrow\downarrow} - U_{\uparrow\uparrow})}
{U_{\uparrow\uparrow} + U_{\downarrow\downarrow} + 2U_{\uparrow\downarrow}}
\le 
\frac{4\alpha_x^2}{mn_0} \left(\frac{\alpha_x^2}{\alpha_z^2} - 1\right)
\label{eqn:stab}
\end{align}
is satisfied. For instance this condition reduces to
$
3U_{\downarrow\downarrow} \ge U_{\uparrow\uparrow} \ge U_{\downarrow\downarrow}/3
$
when $\alpha_z = \alpha_x$ in the $U_{\uparrow\downarrow}\to 0$ limit, revealing 
a peculiar constraint on the strength of the interactions. Our calculation suggests 
that the physical origin of this instability is related to the quantum geometry of the 
underlying space without a deeper insight. 
In addition, when $\alpha_z \ne 0$, Eq.~(\ref{eqn:elow}) further suggests that there is 
a dynamical instability (i.e., $\epsilon_{s\mathbf{q}}$ becomes complex) unless the 
quadratic terms within the square root are positive, i.e.,
$
1 - \alpha_z^2/\alpha_x^2 + mn_0\alpha_z^2(U_{\uparrow\uparrow}
+ U_{\downarrow\downarrow}-2U_{\uparrow\downarrow})/(4\alpha_x^4 \ge 0.
$
This condition is most restrictive when $\alpha_z \to \alpha_x$, giving rise to
$
(U_{\uparrow\uparrow} + U_{\downarrow\downarrow})/2 \ge U_{\uparrow\downarrow}
$
for the dynamical stability of the system. Next we show that the dynamical instability 
never takes place because it is preceded by the so-called roton instability, given that
the geometric mean of $U_{\uparrow\uparrow}$ and $U_{\downarrow\downarrow}$ 
is guaranteed to be less than or equal to the arithmetic mean.

\subsection{Roton Instability at $\mathbf{q} = \mp 2\mathbf{k_0}$}
\label{sec:sgs}

The zero-energy Bogoliubov mode that is found at $\mathbf{q = 0}$ is a special example
of the Goldstone mode that is associated with the spontaneous breaking of
a continuous symmetry in SF systems. In addition to this phonon mode, the Bogoliubov 
spectrum also exhibits the so-called roton mode at finite $\mathbf{q}$. This peculiar 
spectrum clearly originates from the long-range interaction characterized by Eq.~(\ref{eqn:fk}),
and it is a remarkable feature given the surge of recent interest in roton-like spectra in 
various other cold-atom contexts~\cite{santos03, saccani12, macri13, schmidt21, bottcher21}
that paved the way for the creation of dipolar Bose supersolids~\cite{schmidt21, bottcher21}. 
Furthermore the roton spectrum~\cite{khame14, ji15} along with some supersolid 
properties~\cite{li17, putra20} have also been measured with Raman SOC. 
As a consequence of these outstanding progress, the roton spectrum is nowadays 
considered as a possible route and precursor to the solidification of Bose SFs.

Depending on the interaction parameters, our numerics show that there may appear an 
additional pair of zero-energy modes at finite $\mathbf{q}$ when the roton gap vanishes.
See also Refs.~\cite{barnett12, ozawa12, baym14, zhang16a, li13, li15} for 
related observations. 
It turns out they always appear precisely at opposite momentum $\mathbf{q} = \mp 2\mathbf{k_0}$
when the local minimum (maximum) of $\epsilon_{\pm, \mathbf{q}}$ touches 
the zero-energy axis with a quadratic dispersion away from it. 
For instance the roton minimum and its gap is clearly visible in Fig.~\ref{fig:fig1} 
at $q_x = \mp 2\alpha_x$.

Given this numerical observation, we evaluate Eqs.~(\ref{eqn:hpp}) and~(\ref{eqn:hph})
at $\mathbf{q} = \mp 2\mathbf{k_0}$, leading to, e.g., the quasiparticle-quasiparticle element
$
h^{pp}_{-2\mathbf{k_0}} = (U_{\uparrow \uparrow}+U_{\downarrow \downarrow} 
- 2U_{\uparrow \downarrow})n_0/4,
$
quasihole-quasihole element
$
h^{hh}_{-2\mathbf{k_0}} = \varepsilon_{3\mathbf{k_0}} + (U_{\uparrow \uparrow}
+U_{\downarrow \downarrow} + 2U_{\uparrow \downarrow})n_0/4,
$
and quasiparticle-quasihole element
$
h^{ph}_{-2\mathbf{k_0}} = (U_{\downarrow \downarrow} - U_{\uparrow \uparrow})n_0/4.
$
Then, by plugging them into Eq.~(\ref{eqn:esq}), and noting that the stability of the
Bogoliubov theory requires the local minimum (maximum) of the quasiparticle 
(quasihole) spectrum to satisfy
$
\epsilon_{\pm, \mp 2\mathbf{k_0}} \gtrless 0,
$
we obtain the following condition
\begin{align}
\left( \frac{2\alpha_x^2}{mn_0} + U_{\uparrow \uparrow} \right)
\left( \frac{2\alpha_x^2}{mn_0} + U_{\downarrow \downarrow} \right)
> 
\left( \frac{2\alpha_x^2}{mn_0} + U_{\uparrow \downarrow} \right)^2.
\label{eqn:pt}
\end{align}
This condition guarantees the energetic stability of the many-body ground state that is 
presumed in Sec.~\ref{sec:hb} to begin with, and it is in full agreement with 
the previously known results. For instance it reduces to
$
\sqrt{U_{\uparrow \uparrow} U_{\downarrow \downarrow} } > U_{\uparrow \downarrow}
$
in the absence of a SOC when $\alpha_x = 0$, and it reduces to 
$
U > U_{\uparrow \downarrow}
$
for equal intraspin interactions
$
U_{\uparrow \uparrow} = U_{\downarrow \downarrow} = U
$
when $\alpha_x \ne 0$~\cite{cui13, wang10}.
In general Eq.~(\ref{eqn:pt}) suggests that while the ground state is energetically stable 
for all $\alpha_x$ values when 
$
\sqrt{U_{\uparrow \uparrow} U_{\downarrow \downarrow} } > U_{\uparrow \downarrow},
$
it is stable for sufficiently strong SOC strengths $\alpha_x > \alpha_c$ when 
$
\sqrt{U_{\uparrow \uparrow} U_{\downarrow \downarrow} } 
< U_{\uparrow \downarrow}
< (U_{\uparrow \uparrow} + U_{\downarrow \downarrow})/2.
$
Here
$
\alpha_c = [m n_0 
(U_{\uparrow \downarrow}^2 - U_{\uparrow \uparrow} U_{\downarrow \downarrow})
/ (2U_{\uparrow \uparrow} + 2U_{\downarrow \downarrow} - 4U_{\uparrow \downarrow})]^{1/2}
$ 
is the critical threshold.

Both the appearance of an additional pair of zero-energy modes at 
$\mathbf{q} = \mp 2\mathbf{k_0}$ and the associated instability of the many-body 
ground state that is caused by 
$
\epsilon_{\pm, \mathbf{q}} \lessgtr 0
$ 
can be traced back to the degeneracy of the lower-helicity band $\xi_{-,\mathbf{k}}$ 
that is discussed in Sec.~\ref{sec:hb}. For instance, when $\alpha_x \ge \{\alpha_y, \alpha_z\}$, 
our single-particle ground state $|\phi_\mathbf{k_0} \rangle$ is at least two-fold degenerate 
with the opposite-momentum state $|\phi_{-\mathbf{k_0}} \rangle$. Note that the relative 
momentum between these two particle (hole) states is exactly $\mp 2\mathbf{k_0}$. 
Then Eq.~(\ref{eqn:pt}) suggests that while our initial choice for a plane-wave condensate 
that is described purely by the state $|\phi_\mathbf{k_0} \rangle$ is energetically stable 
for sufficiently weak $U_{\uparrow \downarrow}$, it eventually becomes unstable against 
competing states with increasing $U_{\uparrow \downarrow}$. 
Since this instability also occurs precisely at $\mathbf{q} = \mp 2\mathbf{k_0}$, 
it clearly signals the possibility of an additional condensate that is described by the 
state $|\phi_{-\mathbf{k_0}} \rangle$. Thus, when Eq.~(\ref{eqn:pt}) is not satisfied, 
we conclude that the many-body ground state corresponds to the so-called stripe phase 
that is described by a superposition of two states with opposite momentum, 
i.e., $|\phi_\mathbf{k_0} \rangle$ and $|\phi_{-\mathbf{k_0}} \rangle$~\cite{wang10, 
barnett12, li13, li15, zhang16a,zhai15}.
Indeed some supersolid properties of the stripe phase have already been observed
with Raman SOC~\cite{li17, putra20}.

We would like to emphasize that this conclusion is immune to the increased degeneracy 
of the helicity states when the SOC field is isotropic in momentum space. 
For instance, despite the circular degeneracy caused by a Rashba SOC when
$\alpha_x = \alpha_y$, the zero-energy modes still appear at 
$\mathbf{q} = \mp 2\mathbf{k_0}$, and therefore, the stripe phase again involves a 
superposition of two states with opposite momentum.

\subsection{Spurious Jumps at $\mathbf{q} = \mp \mathbf{k_0}$}
\label{sec:jump}
As shown in Fig.~\ref{fig:fig1}, there is an almost perfect agreement between the 
Bogoliubov spectrum of the $4\times4$ Hamiltonian and that of the $2\times2$ 
projected one except for a tiny region in the vicinity of a peculiar jump at 
$\mathbf{q} = \mp \mathbf{k_0}$.
In order to reveal its physical origin, here we set $\alpha_z = 0$ for its
simplicity, and expand the Hamiltonian matrix at 
$\mathbf{q} = -\mathbf{k_0} + \boldsymbol{\delta}$ 
for a small $\boldsymbol{\delta} = (\delta, 0, 0)$. We find that
\begin{align*}
h^{pp}_{\boldsymbol{\delta}} &= \xi_{-, \boldsymbol{\delta}} + \frac{n_0}{4}
\big[ U_{\uparrow\uparrow} + U_{\downarrow\downarrow}
+2U_{\uparrow\downarrow}\cos(\varphi_{\boldsymbol{\delta}}) \big],
\\
h^{ph}_{\boldsymbol{\delta}} &= \frac{n_0}{4}
\big[ U_{\uparrow\uparrow} - U_{\downarrow\downarrow}
+2U_{\uparrow\downarrow}\cos(\varphi_{\boldsymbol{\delta}}) \big],
\end{align*}
where the phase angle $\varphi_{\mathbf{k}}$ is defined in Sec.\ref{sec:hb} leading
to $\cos(\varphi_{\boldsymbol{\delta}}) = \textrm{sgn}(\delta)$. This analysis shows it 
is those coupling terms
$
U_{\uparrow\downarrow}\cos(\varphi_{\boldsymbol{\delta}})
$
between the $\uparrow$ and $\downarrow$ sectors in the Bogoliubov Hamiltonian 
that is responsible for the spurious jump at $\delta = 0$ upon the change of sign
of $\delta$. Note that our initial motivation in deriving the projected Hamiltonian 
in Sec.~\ref{sec:bs} is the assumption that the energy gap between the lower and 
upper helicity bands nearby the single-particle ground state $| \phi_\mathbf{k_0} \rangle$ 
is much larger than the interaction energy. While the validity region of this assumption 
in $\mathbf{k}$ space is not limited with the ground state, it clearly breaks down in 
the vicinity of $\mathbf{k} = \mathbf{0}$ where the $s = \pm$ helicity bands are 
degenerate (see Fig.~\ref{fig:hb}). 
For this reason our projected Hamiltonian becomes unphysical and fails to capture 
the actual result in a tiny region around $\mathbf{q} = -\mathbf{k_0}$.

Having presented a detailed analysis of the Bogoliubov spectrum, next we determine 
the SF density tensor and compare it to the condensate density of the system.

\section{Superfluid $\boldsymbol{\mathrm{vs.}}$ Condensate Density}
\label{sec:scd}

In this paper we define the SF density $\rho_s$ by imposing a so-called phase 
twist on the mean-field order parameter~\cite{taylor06, he12, chen18}. When the SF flows 
uniformly with the momentum $\mathbf{Q}$, the SF order parameter transforms as 
$
\Delta_\sigma \to \Delta_\sigma e^{i \mathbf{Q} \cdot \mathbf{r}},
$
and the SF density tensor $\rho_{ij}$ is defined as the response of the thermodynamic
potential $\Omega_\mathbf{Q}$ to an infinitesimal flow, i.e.,  
\begin{align}
\rho_{ij} = \frac{m}{V} \lim_{\mathbf{Q \to 0}} 
\frac{\partial^2 \Omega_\mathbf{Q}}{\partial Q_i \partial Q_j}.
\end{align}
Here the derivatives are taken for a constant $\Delta_\sigma$ and $\mu_\sigma$, i.e.,
the mean-field parameters do not depend on $\mathbf{Q}$ in the $\mathbf{Q \to 0}$ limit. 
We note that the SF mass density tensor $m \rho_{ij}$ is a related quantity, and it 
corresponds to the total mass involved in the flow.

Let us now calculate $\Omega_\mathbf{Q}$ in the low-$\mathbf{Q}$ limit.
In the absence of an SF flow when $\mathbf{Q = 0}$, the thermodynamic potential 
$\Omega_\mathbf{0}$ can be written as
$
\Omega_\mathbf{0} = \Omega_{zp} + (T/2) \sum_{\ell \mathbf{q}}^\prime 
\mathrm{Tr} \ln \mathbf{G}_{\mathbf{0} \ell \mathbf{q}}^{-1},
$
where
$
\Omega_{zp} = -\mu N_0/2 - \sum_\mathbf{q}^\prime  \big( \varepsilon_\mathbf{q} + \mu \big)/2
$
is the zero-point contribution, $T$ is the temperature with the Boltzmann 
constant $k_\mathrm{B} = 1$, $\mathrm{Tr}$ is the trace, and 
$
\mathbf{G}_{\mathbf{0} \ell \mathbf{q}}^{-1}= i\omega_\ell \sigma_0 \tau_z - \mathbf{H}_\mathbf{q}
$
is the inverse of the Green's function for the Bogoliubov Hamiltonian that is given 
in Eq.~(\ref{eqn:HB}). 
Here $\omega_\ell = 2 \pi \ell T$ is the bosonic Matsubara frequency with 
$\ell$ an integer. In order to make some analytical progress, we make use of the
Bogoliubov states and spectrum determined by Eq.~(\ref{eqn:bogmod}), 
and define~\cite{julku21a, julku21b}
\begin{align}
\mathbf{G}_{\mathbf{0} \ell \mathbf{q}} = \sum_{n s}
\frac{s|\chi_{s \mathbf{q}}^n \rangle \langle \chi_{s \mathbf{q}}^n |} 
{i\omega_\ell - E_{s \mathbf{q}}^n}.
\end{align}
This expression clearly satisfies 
$
\mathbf{G}_{\mathbf{0} \ell \mathbf{q}}^{-1} \mathbf{G}_{\mathbf{0} \ell \mathbf{q}} = \sigma_0 \tau_0.
$
In the presence of an SF flow when $\mathbf{Q \ne 0}$, the thermodynamic 
potential $\Omega_\mathbf{Q}$ can be obtained through a gauge transformation 
of the bosonic field operators
$
\tilde{a}_{\sigma \mathbf{q}} \to \tilde{a}_{\sigma \mathbf{q}} e^{i \mathbf{Q} \cdot \mathbf{r}}.
$
This transformation removes the phases of the SF order parameters, and we obtain
the inverse Green's function
$
\mathbf{G}_{\mathbf{Q} \ell \mathbf{q}}^{-1} =  \mathbf{G}_{\mathbf{0} \ell \mathbf{q}}^{-1}  
- \boldsymbol{\Sigma}_\mathbf{Q}
$
of the twisted system. Its $\mathbf{Q}$-dependent part has three terms
$
\boldsymbol{\Sigma}_\mathbf{Q} = \boldsymbol{\Sigma}_{\mathbf{Q}, 1} 
+ \boldsymbol{\Sigma}_{\mathbf{Q}, 2} + \boldsymbol{\Sigma}_{\mathbf{Q}, 3}
$
~\cite{he12}: while the SOC-independent terms
$
\boldsymbol{\Sigma}_{\mathbf{Q}, 1} = \frac{Q^2}{2m} \sigma_0 \tau_0
$
and
$
\boldsymbol{\Sigma}_{\mathbf{Q},2} = \frac{\sigma_0}{m}
\begin{bmatrix}
(\mathbf{k_0+q}) \cdot \mathbf{Q} & 0 \\
0 & (\mathbf{k_0-q}) \cdot \mathbf{Q}
\end{bmatrix}
$
are diagonal both in the spin and particle-hole sectors, the SOC-induced term
$
\boldsymbol{\Sigma}_{\mathbf{Q},3} = \frac{1}{m}
\begin{pmatrix}
\mathbf{d}_\mathbf{Q} \cdot \boldsymbol{\sigma} & 0 \\
0 & \mathbf{d}_\mathbf{Q} \cdot \boldsymbol{\sigma}^*
\end{pmatrix}
$
is diagonal only in the particle-hole sector. These terms can be conveniently reexpressed as
$
\boldsymbol{\Sigma}_{\mathbf{Q},1} = (1/2) \sum_{ij} Q_i Q_j \partial_i \partial_j \mathbf{H}_\mathbf{q}
$
and
$
\boldsymbol{\Sigma}_{\mathbf{Q},2+3} = \sum_i Q_i \tau_z \partial_i \mathbf{H}_\mathbf{q},
$
where $\partial_i \mathbf{H}_\mathbf{q}$ stands for 
$
\partial \mathbf{H}_\mathbf{q}/\partial q_i.
$

\begin{figure*}[!htb]
\centering
\resizebox{2.0\columnwidth}{!}{%
\includegraphics{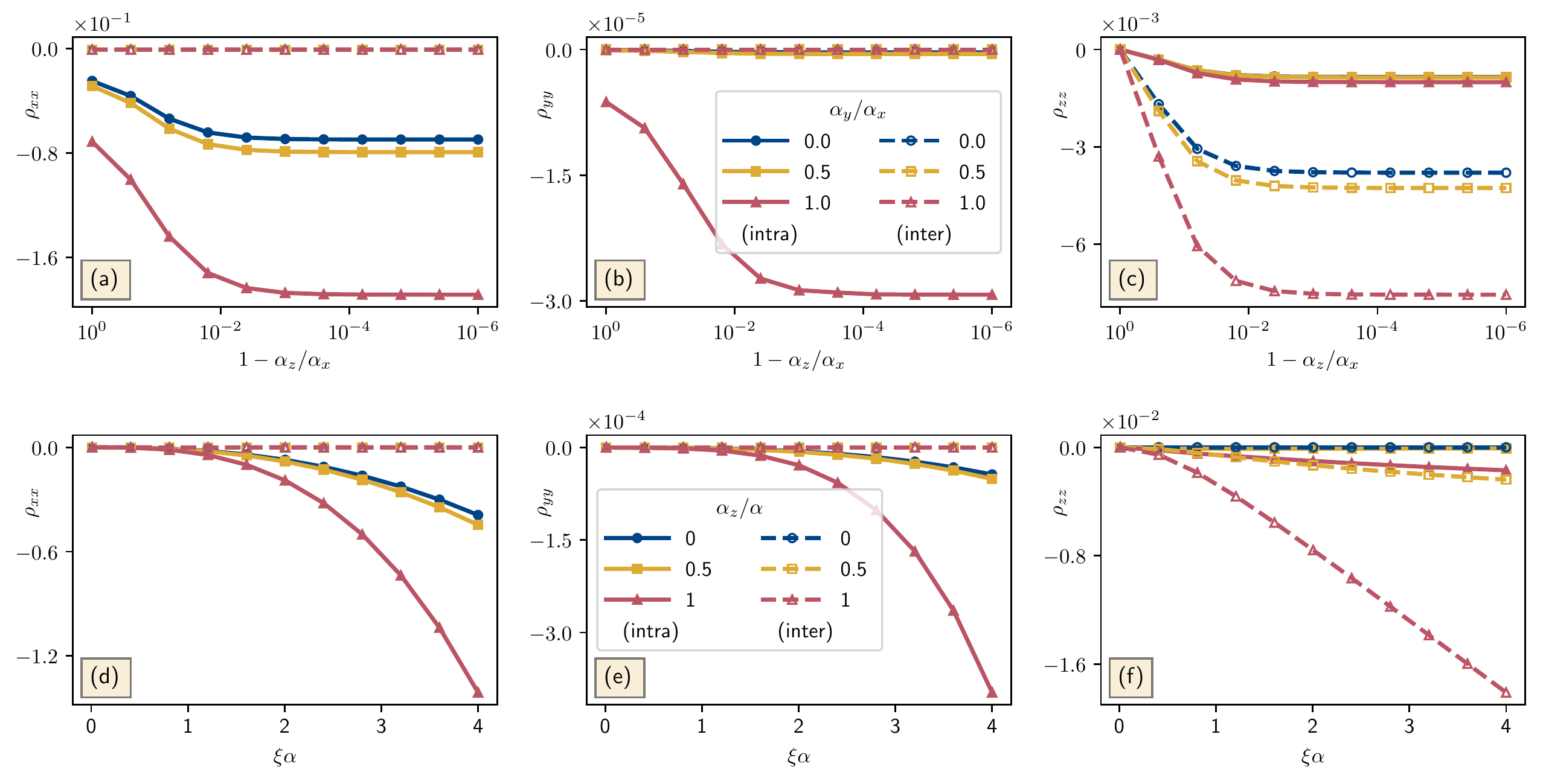}}
\caption{The intraband (solid lines) and interband (dashed lines) contributions to the summation 
term in the superfluid-density tensor $\rho_{ij}$ are shown when $T = 0$ and
$
U = U_{\uparrow\uparrow}= U_{\downarrow\downarrow} = 10 U_{\uparrow\downarrow} /9.
$
Here the left, middle and right columns correspond, respectively, to the diagonal elements 
$\rho_{xx}$, $\rho_{yy}$ and $\rho_{zz}$ (in units of $n_0 \approx n$), and all of the 
off-diagonal elements vanish.
In the top panel the diagonal elements are shown as a function of $\alpha_z$ for three 
values of $\alpha_y$ when $\alpha_x = 2/\xi$ is fixed. 
In the bottom panel the diagonal elements are shown as a function of the SOC strength 
$\alpha = \alpha_x = \alpha_y$ for three values of the $\alpha_z/\alpha$ ratio.
\label{fig:sd}}
\end{figure*}

Since we are interested only in the low-$\mathbf{Q}$ limit of $\Omega_\mathbf{Q}$,
we can use the Taylor expansion
$
\ln \det \mathbf{G}_{\mathbf{Q} \ell \mathbf{q}}^{-1} = 
\mathrm{Tr} \ln \mathbf{G}_{\mathbf{0} \ell \mathbf{q}}^{-1} - 
\mathrm{Tr} \sum_{l = 1}^\infty (\mathbf{G}_{\mathbf{0} \ell \mathbf{q}} \mathbf{\Sigma}_\mathbf{Q})^l/l,
$
and keep up to second-order terms in $\mathbf{\Sigma}_\mathbf{Q}$. 
This calculation leads to
\begin{align}
\rho_{ij} = & m n_0 M^{-1}_{ij} - \frac{m T}{2V} \sum_{\ell \mathbf{q}}^\prime 
\big[
\mathrm{Tr} \big(
\mathbf{G}_{\mathbf{0} \ell \mathbf{q}} \partial_i \partial_j \mathbf{H}_\mathbf{q} \big)
\nonumber \\
&+\mathrm{Tr} \big(
\mathbf{G}_{\mathbf{0} \ell \mathbf{q}} \tau_z \partial_i \mathbf{H}_\mathbf{q}
\mathbf{G}_{\mathbf{0} \ell \mathbf{q}} \tau_z \partial_j \mathbf{H}_\mathbf{q}
\big) \big].
\end{align}
Here the first term is due to the kinetic energy of the condensate in the presence 
of an SF flow, i.e., there is an additional quadratic contribution $(N_0/2) \sum_{ij} Q_i Q_jM^{-1}_{ij}$ 
to $\Omega_{zp}$ coming from the low-$\mathbf{Q}$ expansion of
$
\sum_\mathbf{k} \xi_{-, \mathbf{k+Q}}
a_{-, \mathbf{k}}^\dagger a_{-, \mathbf{k}}
$
around $\mathbf{k_0}$.
Thus, when the inverse of the effective-mass tensor $M^{-1}_{ij}$ vanishes, $\rho_{ij}$ is 
determined entirely by the Bogoliubov Hamiltonian, i.e., the quantum fluctuations above 
the condensate. The trace of the Green's function in the second term is related 
to the density of excited (non-condensate) particles $n_e$ since its diagonal 
elements yield
$
n_e = -(T/V) \sum_{\ell \mathbf{q}}^\prime  \big(
G_{\mathbf{0}\ell \mathbf{q}}^{11} + G_{\mathbf{0} \ell \mathbf{q}}^{22}
\big) e^{-i \omega_\ell 0^+},
$
or alternatively,
$
n_e= -(T/V) \sum_{\ell \mathbf{q}}^\prime  \big(
G_{\mathbf{0} \ell \mathbf{q}}^{33} + G_{\mathbf{0} \ell \mathbf{q}}^{44}
\big) e^{i \omega_\ell 0^+}.
$
Thus, by performing the summation over the Matsubara frequencies, we eventually 
obtain
\begin{align}
&\rho_{ij} = n_t \delta_{ij} - n_0\frac{\alpha_y^2 \delta_{iy} + \alpha_z^2\delta_{iz}}{\alpha_x^2} 
+ \frac{m}{2V} \sum_{nn'ss'\mathbf{q}}^\prime  ss'
\langle \chi_{s \mathbf{q}}^n | \tau_z 
\nonumber \\
& \partial_i \mathbf{H}_\mathbf{q} | \chi_{s' \mathbf{q}}^{n'} \rangle 
\langle \chi_{s' \mathbf{q}}^{n'} | \tau_z \partial_j \mathbf{H}_\mathbf{q} | \chi_{s \mathbf{q}}^n \rangle 
\frac{f_\mathrm{B}(E_{s\mathbf{q}}^n) - f_\mathrm{B}(E_{s'\mathbf{q}}^{n'})} 
{E_{s\mathbf{q}}^n - E_{s'\mathbf{q}}^{n'}},
\label{eqn:rhoij}
\end{align}
where $n_t = n_0 + n_e$ is the total density of particles in the system and $f_\mathrm{B}(x)$ 
is the Bose-Einstein distribution function. Here a partial derivative
$
\partial f_\mathrm{B}(E_{s \mathbf{q}}^n) / \partial E_{s \mathbf{q}}^n = 
-[1/(4T)]\mathrm{cosech}^2\left[E_{s \mathbf{q}}^n/(2T)\right]
$
is implied when the summation indices coincide simultaneously ($n = n'$ and $s = s'$). 
In comparison to the SF density, the non-condensate density can be written as
\begin{align}
n_\mathrm{e} &= \frac{1}{2V} \sum_{ns\mathbf{q}}^\prime s
\big[ \langle \chi_{s \mathbf{q}}^n | \chi_{s \mathbf{q}}^n \rangle f_\mathrm{B}(E_{s\mathbf{q}}^n) 
+ {_2\langle} \chi_{s \mathbf{q}}^n | \chi_{s \mathbf{q}}^n \rangle_2
\big], \\
&= \frac{1}{2V} \sum_{ns\mathbf{q}}^\prime 
\big[ s \langle \chi_{s \mathbf{q}}^n | \chi_{s \mathbf{q}}^n \rangle f_\mathrm{B}(E_{s\mathbf{q}}^n) - 1/2
\big].
\label{eqn:ne}
\end{align}
We checked that both expressions yield the same numerical result.
Note that
$
n_\mathrm{e}^0 =  [1/(2V)] \sum_{n \mathbf{q}}^\prime 
( -1 + \langle \chi_{-, \mathbf{q}}^n | \chi_{-, \mathbf{q}}^n \rangle )
$
is the so-called quantum depletion of the condensate at $T = 0$.

As an illustration, in the case of a single-component Bose gas, there is a single 
Bogoliubov band with the usual quasiparticle-quasihole symmetric spectrum
$
E_{s \mathbf{q}} = s E_\mathbf{q}
$
where
$
E_\mathbf{q} = \sqrt{\varepsilon_\mathbf{q}(\varepsilon_\mathbf{q} + 2U n_0)},
$
and by plugging
$
\langle \chi_{s \mathbf{q}} | \tau_z \partial_i \mathbf{H}_\mathbf{q} | \chi_{s' \mathbf{q}} \rangle 
= (s q_i/m) \delta_{s s'}
$
into Eq.~(\ref{eqn:rhoij}), we recover the textbook definition
$
\rho_{ij} = n_t \delta_{ij} + [1/(mV)] \sum_\mathbf{q}^\prime  q_i q_j 
\partial f_\mathrm{B}(E_\mathbf{q}) / \partial E_\mathbf{q}
$
of $\rho_s$~\cite{fetter}. This shows that $\rho_{ij} = n_t \delta_{ij}$ at zero temperature 
and that the entire gas is SF. Similarly, by plugging
$
\langle \chi_{s \mathbf{q}} | \chi_{s' \mathbf{q}} \rangle = (\varepsilon_\mathbf{q} + Un_0)/E_\mathbf{q}
$
into Eq.~(\ref{eqn:ne}), we recover the textbook definition of $n_e = n_e^0 + n_e^T$, where
$
n_e^0 = [1/(2V)] \sum_\mathbf{q}^\prime  [-1 + (\varepsilon_\mathbf{q} + Un_0)/E_\mathbf{q}]
$
is the quantum depletion and
$
n_e^T =  (1/V) \sum_\mathbf{q}^\prime  (\varepsilon_\mathbf{q} + Un_0) 
f_\mathrm{B}(E_\mathbf{q}) / E_\mathbf{q}
$
is the thermal one~\cite{fetter}.

We note in passing that Eq.~(\ref{eqn:rhoij}) is consistent with the so-called SF weight 
that is derived in Ref.~\cite{julku21a, julku21b} for a multi-band Bloch Hamiltonian. 
See also~\cite{yangnote}. 
Unlike our phase-twist method, they define the SF weight as the long-wavelength and 
zero-frequency limit of the current-current linear response.
In particular our expression for a continuum model is formally equivalent to their 
$
D_{1, \mu\nu}^s + D_{2, \mu\nu}^s + D_{3, \mu\nu}^s
$ 
with the caveat that $D_{2, \mu\nu}^s$ is cancelled by the interband contribution of 
$D_{1, \mu\nu}^s$. This is similar to the cancellation that they observed for the 
Kagome lattice. Furthermore Eq.~(\ref{eqn:rhoij}) can also be split into two parts 
$
\rho_{ij} = \rho_{ij}^\mathrm{intra} + \rho_{ij}^\mathrm{inter}
$ 
depending on the physical origin of the terms~\cite{julku21a, julku21b}: the intraband (interband) 
processes give rise to the conventional (geometric) contribution. This division is motivated 
by the success of a similar description with Fermi SFs~\cite{iskin18a, julku18}.

\begin{figure}[!htb]
\centering
\resizebox{.75\columnwidth}{!}{%
\includegraphics{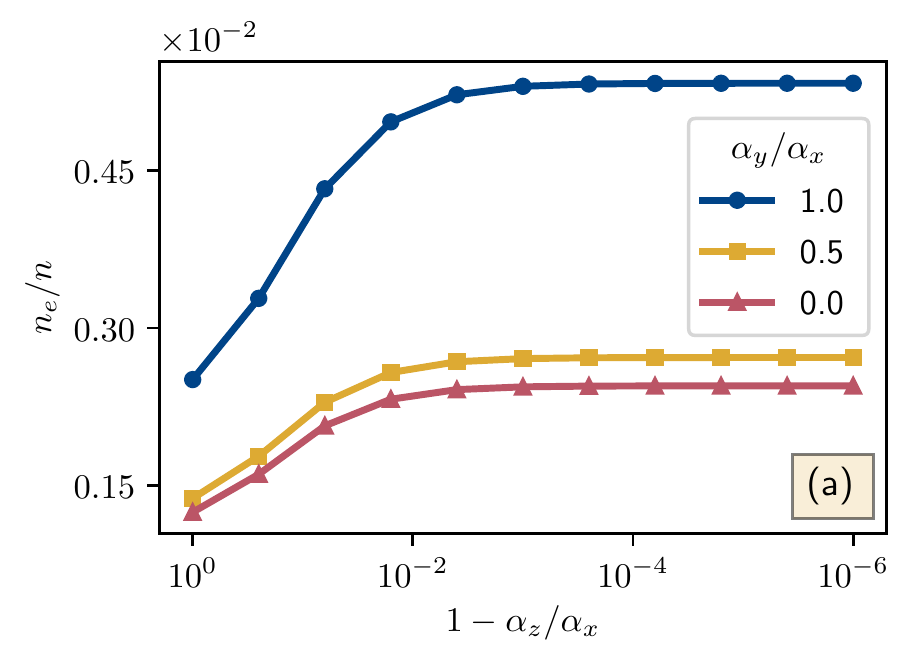}}
\resizebox{.75\columnwidth}{!}{%
\includegraphics{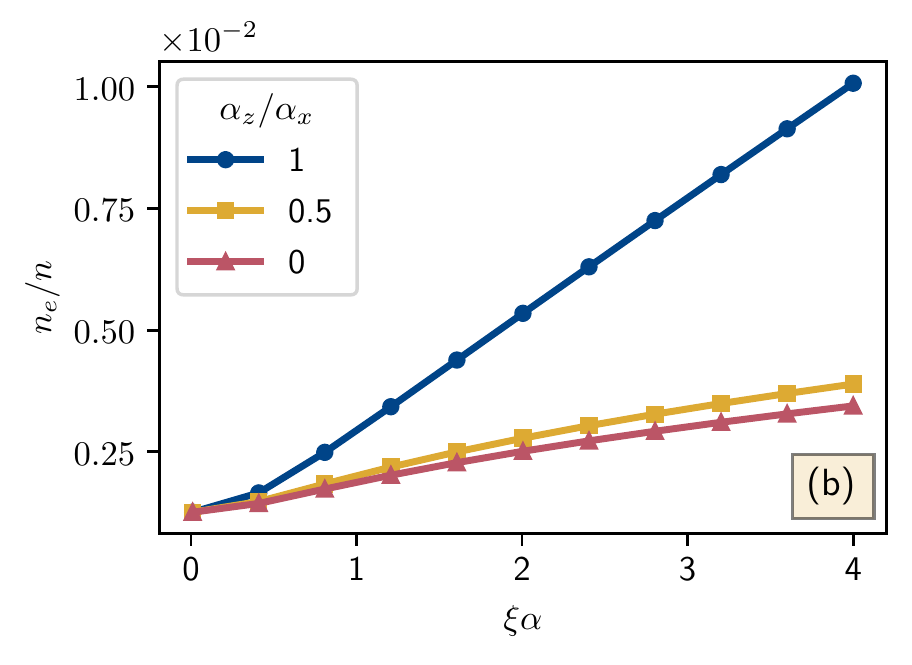}}
\caption{
Quantum depletion of the condensate density is shown as a function of SOC 
parameters when $T = 0$ and
$
U = U_{\uparrow\uparrow}= U_{\downarrow\downarrow} = 10 U_{\uparrow\downarrow} / 9.
$ 
The fraction of the depletion is plotted (a) as a function of $\alpha_z$ for 
three values of $\alpha_y$ when $\alpha_x = 2/\xi$ is fixed, and
(b) as a function of the SOC strength $\alpha = \alpha_x = \alpha_y$ for three values 
of the $\alpha_z/\alpha$ ratio. 
\label{fig:depletion}}
\end{figure}

In order to provide further evidence for its geometric origin, in Fig.~\ref{fig:sd} we 
compare the interband contribution with that of the intraband one coming from the 
summation term in Eq.~(\ref{eqn:rhoij}). Here we set $T \to 0$.
First of all this figure shows that the total contribution from the summation 
term decreases with the increased strength and isotropy of the SOC fields, 
i.e., when $\alpha_y \to \alpha_x$ in Figs.~\ref{fig:sd}(a,b,c) and when 
$\alpha_z \to \alpha$ in Figs.~\ref{fig:depletion}(d,e,f). Thus $\rho_{xx}$ always 
decreases from $n_t$ with SOC. However, depending on the value of $\alpha_y$ 
and $\alpha_z$, the remaining contribution 
$
n_e + n_0(\alpha_x^2 - \alpha_y^2 \delta_{iy} - \alpha_z^2\delta_{iz}) / \alpha_x^2
$
to $\rho_{yy}$ and $\rho_{zz}$ in Eq.~(\ref{eqn:rhoij}) may compete with or favor the 
contribution from the summation term. More importantly Fig.~\ref{fig:sd} shows that 
not only $\rho_{zz}$ has the largest interband contribution but also its relative 
weight is predominantly controlled by $\alpha_z \ne 0$.
These findings are in support of our Bogoliubov dispersion given in Eq.~(\ref{eqn:elow}) 
whose quantum-geometric contributions are fully controlled by $\alpha_z \ne 0$.
For completeness, in Fig.~\ref{fig:depletion} we present the quantum depletion 
$n_e^0$ as a function of SOC parameters when 
$
U = U_{\uparrow\uparrow}= U_{\downarrow\downarrow} = 10 U_{\uparrow\downarrow} / 9.
$ 
This figure shows that $n_e^0$ increases with the increased strength and isotropy 
of the SOC fields~\cite{cui13}, i.e., when $\alpha_y \to \alpha_x$ in 
Fig.~\ref{fig:depletion}(a) and when $\alpha_z \to \alpha$ in Fig.~\ref{fig:depletion}(b). 
This is clearly a direct consequence of the increased degeneracy of the single-particle 
spectrum. However, since $n_e^0 \ll n$ even for moderately strong SOC fields, 
the Bogoliubov approximation is expected to work well in general.

\section{Conclusion}
\label{sec:conc}

To summarize here we considered the plane-wave BEC phase of a spin-orbit-coupled 
Bose gas and reexamined its SF properties from a quantum-geometric perspective. 
In order to achieve this task analytically, we first reduce the $4 \times 4$ Bogoliubov
Hamiltonian (that involves both lower and upper helicity bands) down to $2 \times 2$ 
through projecting the system onto the lower helicity band. This is motivated by the 
assumption that the energy gap between the lower and upper helicity bands nearby 
the single-particle ground state is much larger than the interaction energy.
Then, given our numerical verification that the projected Hamiltonian provides an almost 
perfect description for the lower (higher) quasiparticle (quasihole) branch in the 
Bogoliubov spectrum, we exploited the low-momentum Bogoliubov spectrum analytically 
and identified the geometric contributions to the sound velocity. 
In contrast to the conventional contribution that has a square-root 
dependence on the interaction strength, we found that the geometric ones are distinguished 
by a linear dependence. It may be important to emphasize that these geometric effects 
are not caused by the negligence of the upper helicity band.
Similar to the Fermion problem where the geometric effects dress the effective mass 
of the Goldstone modes~\cite{iskin20a, iskin20b}, here one can also interpret 
the geometric terms in terms of a dressed effective mass for the Bogoliubov modes.
We also discussed the roton instability of the plane-wave ground state against the 
stripe phase and determined the phase transition boundary. 
In addition we derived the SF density tensor by imposing a phase-twist on 
the condensate order parameter and analyzed the relative importance of its contribution 
from the interband processes that is related to the quantum geometry. 
As an outlook we believe it is worthwhile to do a similar analysis for the stripe phase.

\begin{acknowledgments}
We thank Aleksi Julku for email correspondence and M. I. acknowledges funding from 
T{\"U}B{\.I}TAK Grant No. 1001-118F359. 
\end{acknowledgments}


\begin{thebibliography}{99}
\bibitem{yangnote}
Towards the completion of this work, we became aware of an independent 
study~\cite{yang21}, where the SF density tensor is derived for a two-dimensional system 
using the linear-response theory. The results seem to agree with ours when there is an overlap.
\end{thebibliography}
\end{document}